\documentclass[final]{aipproc}
\layoutstyle{6x9}

\begin{document}

\title{Non-perturbative description of self-interacting charged scalar field at finite 
temperature and in the presence of an external magnetic field}

\author{D. C. Duarte}{
  address={Departamento de Ci\^encias Naturais, Universidade Federal de S\~ao Jo\~ao del
Rei, 36301-000 S\~ao Jo\~ao del Rei, MG, Brazil} }

\author{R. L. S. Farias}{
  address={Departamento de Ci\^encias Naturais, Universidade Federal de S\~ao Jo\~ao del
Rei, 36301-000 S\~ao Jo\~ao del Rei, MG, Brazil} }

\author{Rudnei O. Ramos}{
  address={Departamento de
  F\'{\i}sica Te\'orica, Universidade do Estado do Rio de Janeiro, 20550-013
  Rio de Janeiro, RJ, Brazil} }

\begin{abstract}

It is studied the symmetry restoration in a theory of a self-interacting charged scalar field at 
finite temperature and in
the presence of an external magnetic field. The effective potential is evaluated 
nonperturbatively through the Optimized Perturbation Theory (OPT) nonperturbative method. 
In addition, we present an efficient numeric way to deal with the sum over the Landau levels 
in the case of working with weak magnetic fields.

\end{abstract}

\maketitle
\vspace{-0.75cm}

Phase transition in quantum field theory are phenomena of typically nonperturbative nature and have been a subject of 
interest due to their range of possible applications, going from low energy phenomena in condensed matter systems to 
high-energy phase transitions in particle physics and cosmology. 
Typical examples where these problems can manifest are in studies of symmetry changing phenomena in a hot and dense medium. 
Consequently, there is a lot of interest in investigating thermal field theories describing matter under extreme 
conditions~\cite{blaizot}. 
In this work, we analyze the phase transition for a self-interacting complex scalar field model
and determine how an external magnetic field, combined with thermal effects, affects the transition. All calculations were 
performed in the context of the OPT nonperturbative method. {}Furthermore, we also present an efficient way to deal with the 
sum over Landau levels based on the Euler-Maclaurin (EM) formula~\cite{ourPRD}, which is
particularly suitable for numerical work and of interest especially in the case of working with weak magnetic fields, 
where a very large number of Landau levels has to be taken in account.

\vspace{-0.75cm}
\section{The Optimized Perturbation Theory}

In this study, we use a self-interacting quartic complex scalar field model with $U(1)$ global
symmetry with spontaneous symmetry breaking at tree level in the potential, whose Lagrangian density
can be written as
\begin{equation}
\mathcal{L} = \partial_{\mu}\phi\partial^{\mu}\phi^* + m_0^2|\phi|^2
-\frac{\lambda}{6}|\phi|^4\;,
\end{equation}
where $m^2 > 0$ for spontaneous symmetry breaking. 
The implementation of the OPT in Lagrangian density is the standard one~\cite{rud2} 
and is implemented through an interpolation procedure, with result
\begin{eqnarray}
 \mathcal{L}\to\mathcal{L_{\delta}} & = & \sum_{i = 1}^{2} \left[\frac{1}{2}\left(
\partial_{\mu}\phi_i\right)^2 - \frac{1}{2}\Omega^2\phi_i^2 + \frac{\delta}{2}\eta^2\phi_i^2
- \frac{\delta\lambda}{4!}(\phi_i^2)^2\right]\nonumber\\
 & + & \Delta\mathcal{L}_{\rm{ct},\delta}\;,
\end{eqnarray}
where the complex field $\phi$ is written in terms of real and imaginary components, $\phi = \frac{1}{\sqrt{2}}(\phi_1 + i\phi_2)$,
the mass term $\Omega^2 = -m^2 + \eta^2$ and $\Delta\mathcal{L}_{\rm{ct},\delta}$ is the Lagrangian density
part with the renormalization counterterms needed to render the theory finite. The $\delta$ expansion 
parameter is set equal to one at the end and $\eta$ is an arbitrary mass parameter.
The effective potential, in terms of {}Feynman diagrams, in the OPT formalism
at $\mathcal{O}\left(\delta^1\right)$ reads
\vspace{-0.3cm}
\begin{eqnarray*}
\hspace{-0.45 cm}
\begin{picture}(400,5) \put(70,-1){$V_{eff}\left( \phi \right)=
V_0\left(\phi\right) +$} \thicklines \put(180,0){\circle{20}} \put(200,-1) {+}
\put(225,0){\circle{20}}\put(245,0){\circle{20}} \put(265,-1)
{+} \put(290,0){\circle{20}}\put(300,0){\circle*{5}}
\put(315,-1) {+}
\put(395,0)\thicklines \put(340,0){\line(-1,-1){10}}
\put(350,0){\circle{20}} \put(340,0){\line(-1,1){10}}
\end{picture}
\end{eqnarray*}

The OPT with optimization of the effective potential implements
automatically a nonperturbative resummation of the thermal corrections, 
with the Principle of Minimal Sensitivity (PMS)~\cite{prd1-lde}, defined by
\begin{equation}
 \left.\frac{\partial V_{\rm eff}(\eta)}{\partial \eta}\right|_{\bar{\eta}}=0\;.
\label{pms}
\end{equation}
The optimum value $\eta$ is a function of the original parameters of the theory. In particular,
$\eta$ is a nontrivial function of the couplings and because of this, nonperturbative results
are generated when the optimization procedure is applied directly to the effective potential.



A serious numerical problem arises in a system where the phase structure behavior in weak magnetic field limit is studied: 
It is necessary to take into account a very large number of Landau levels to make the sum converge. To deal with this problem, we use 
successfully the EM formula that provides a connection between integrals and sums and can be used to evaluate finite sums 
and series using integrals or vice-versa. Its general form is
\vspace{-0.2cm}
\begin{eqnarray}
 \sum_{k = a}^b f(k) & = & \int_a^b f(x)dx + \frac{1}{2}\left[f(a) - f(b)\right] + \sum_{i=1}^n \frac{b_{2i}}{(2i)!}\left[f^{(2i-1)}(b) 
 - f^{(2i-1)}(a)\right] \nonumber\\
& + & \int_a^b\frac{B_{2n + 1}(\{x\})}{(2n + 1)!} f^{(2n+1)(x)dx}\;,
\label{EMform}
\end{eqnarray}
\vspace{-0.05cm}
where $b_i$ are the Bernoulli numbers, and $B_n(x)$ are the Bernoulli polynomials. The notation $\{x\}$
in Eq.~(\ref{EMform}) means the fractional part of $x$ and $f^{(k)}(x)$ means the $k$th derivative of the function.
The last term in Eq.~(\ref{EMform}) is known as the \emph{remainder}.
The EM approximation produces results that already at first order in $n$ have good accuracy when compared to the 
exact values coming from the Landau sum and the remainder is taken into account. It is important to note that Eq.~(\ref{EMform}) is not 
an approximation to the sum, but it is actually an identity.

\vspace{-0.9cm}

\section{Numerical Results}

\vspace{-0.2cm}

With the application of the optimization procedure (PMS) given by Eq.~(\ref{pms}) in the OPT context
we obtain the numerical results in the figures shown below, where all quantities are in units of the 
arbitrary mass regularization scale $M$:

\vspace{0.29cm}

\begin{figure}[ht]
\centerline{
\includegraphics[height=3.5cm]{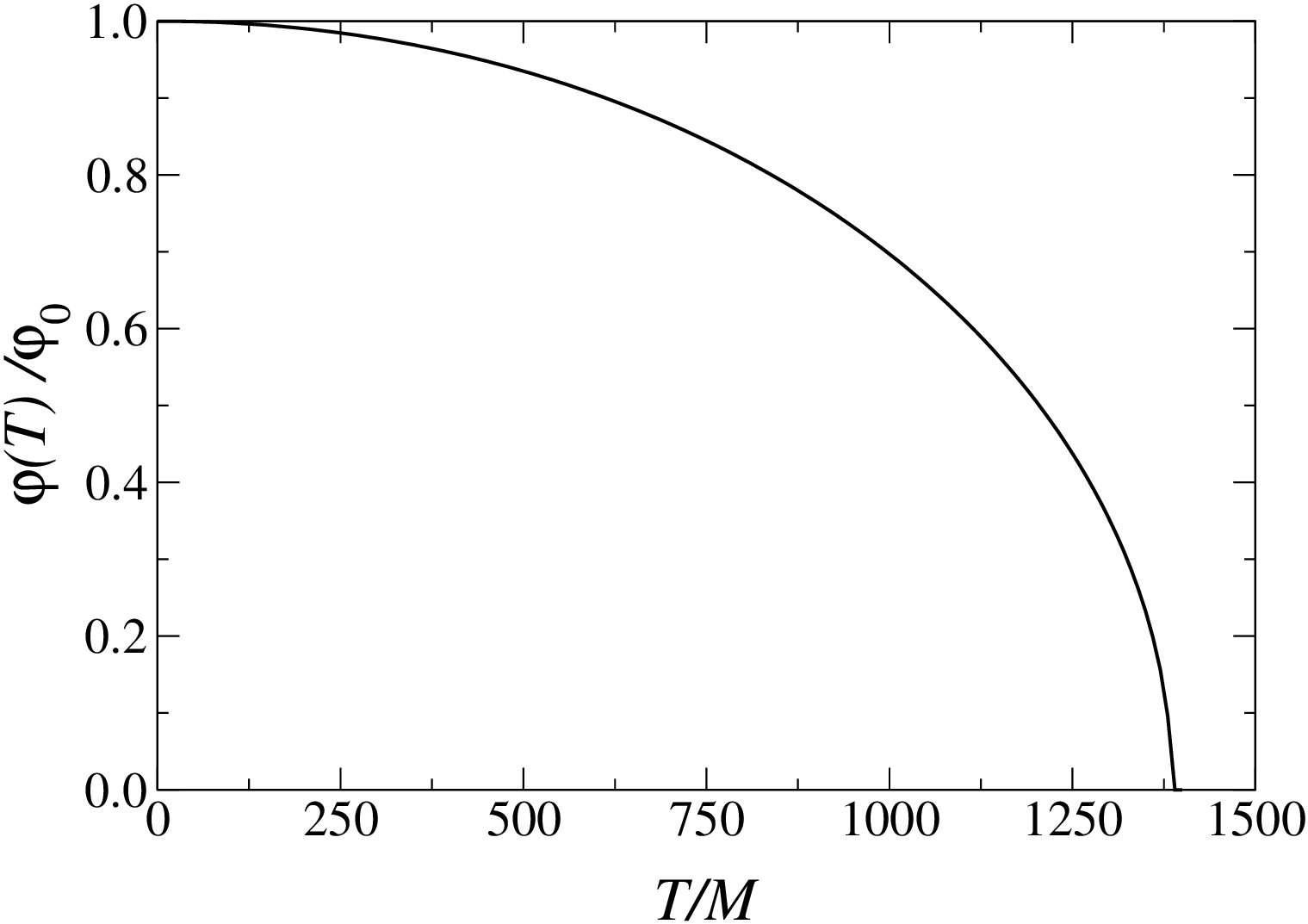}
\includegraphics[height=3.5cm]{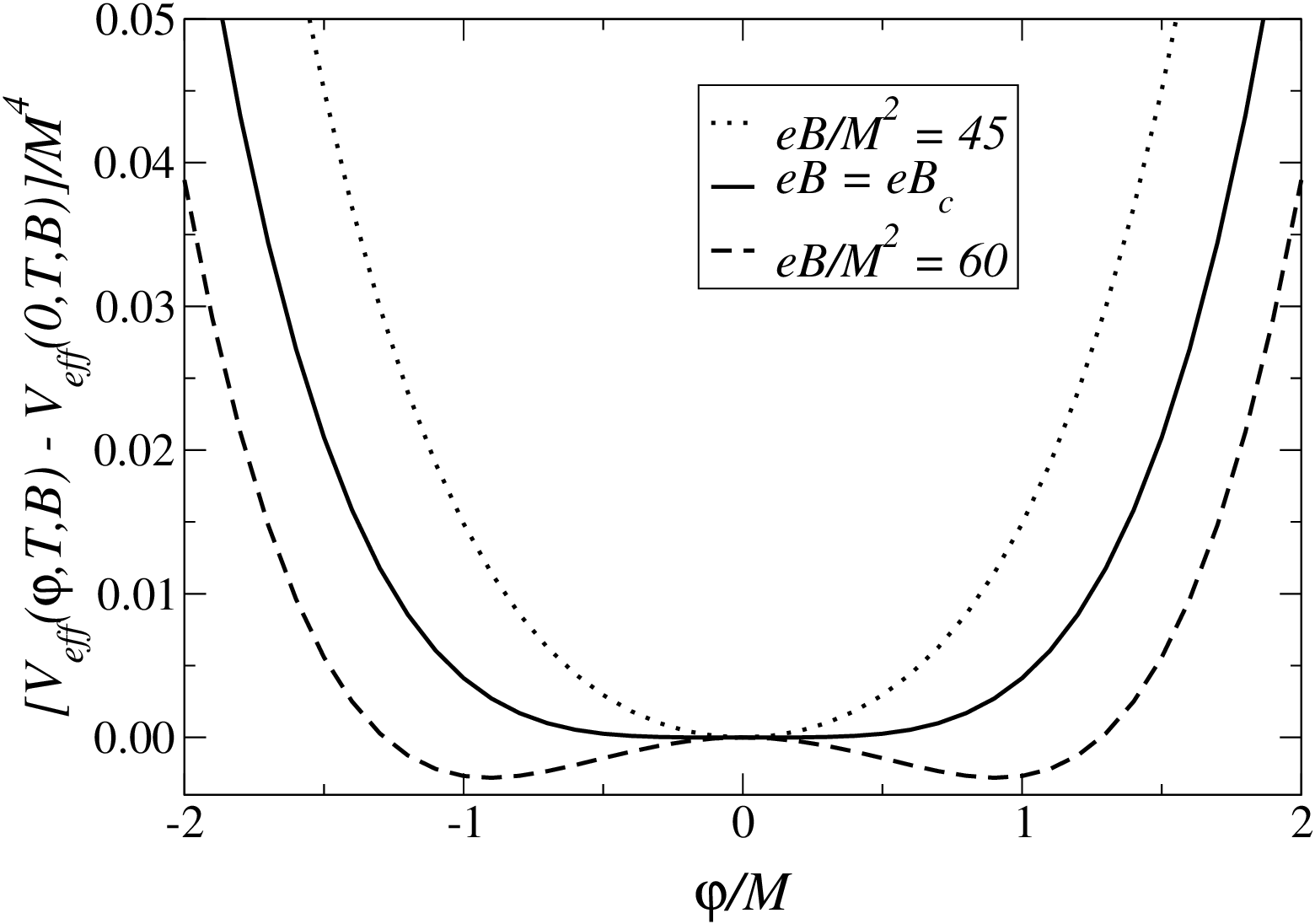}}
\end{figure}
\vspace{0.05cm}
\begin{figure}[ht]
 \centerline{
\includegraphics[height=3.5cm]{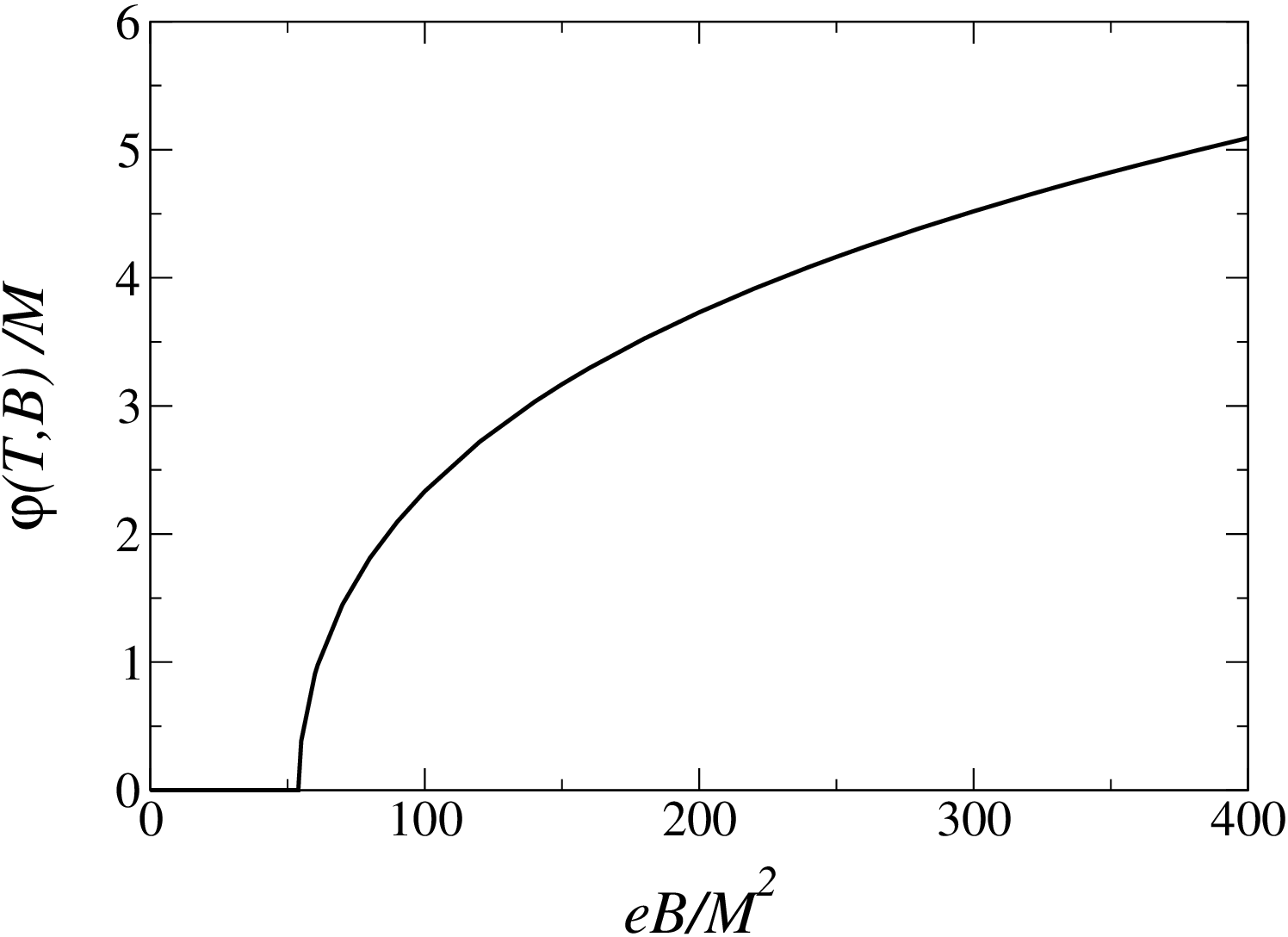}
\hspace{0.2cm}
\includegraphics[height=3.5cm]{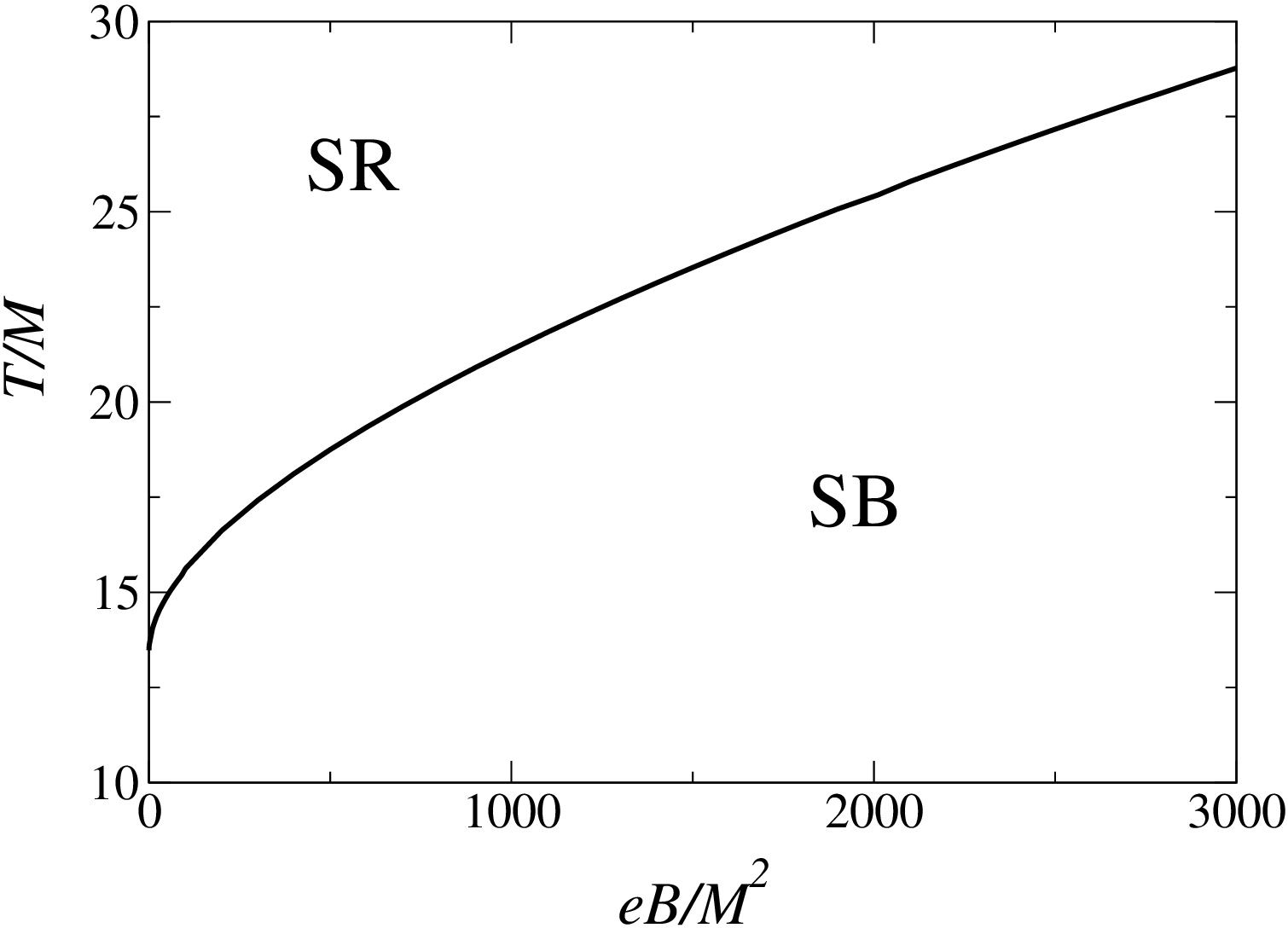}}
\caption{\textbf{First line}: Temperature dependence of the VEV, where $\varphi_0$ is the minimum of the tree-level potential for $B=0$.
Parameters: $m/M = 20$, $\lambda = 0.00375$ (left) and effective potential for three different values of $B$. 
Parameters: $m/M = 1, \lambda = 0.1, T/M = 15$ and $eB_c/M^2 \simeq 53.93$ (right).
 \textbf{Second line}: Magnetic field dependence of the VEV for a fixed $T$ above $T_c$. Parameters: $m/M = 1, \lambda = 0.1$
and $T/M = 15$ (left) and phase diagram of the system on $(B,T)$ plane. Parameters: 
$m/M = 1, \lambda = 0.1$ (right). }
\label{results}
\end{figure}
\vspace{-0.2cm}
The renormalization procedure is not affected by the introduction of an external magnetic field effects.
The counterterms required to renormalize the theory are the same in the case of $T = 0$. In both limits, for strong and weak magnetic 
fields (according to the value of the ratio $eB/T^2$) a second order phase transition is obtained. This is in disagreement with 
results obtained using a different resummation method, such as in~\cite{ayala}. 
The effect of the external magnetic field is to strengthen the symmetry-broken phase, delaying the phase transition. 

Our results for the variation of critical temperature with the magnetic field, already using the EM formula, 
are in agreement with different theoretical models~\cite{andersen}, 
despite being in disagreement with lattice results.

\vspace{-0.5cm}
\begin{theacknowledgments}
Work partially financed by CNPq and Faperj.
\end{theacknowledgments}
\vspace{-0.5cm}

\bibliographystyle{aipproc}   
\bibliographystyle{aipprocl} 

\bibliography{HADRONSproc}

\end{document}